\documentclass[twocolumn,9pt]{article}

\usepackage{balance}
\usepackage{etoolbox}
\usepackage[export]{adjustbox}
\usepackage[utf8]{inputenc}
\usepackage{flushend}
\usepackage[section]{placeins}
\usepackage[bf, footnotesize]{caption}
\usepackage{graphics}
\usepackage{CJKutf8}

\usepackage{amssymb}
\usepackage{mathtools, cuted}
\usepackage{widetext}

\usepackage{lineno}

\usepackage{multicol}
\usepackage{booktabs}
\usepackage{amsmath}
\usepackage{widetext}
\usepackage{flushend}
\usepackage[explicit]{titlesec}
\titleformat{\section}
  {\normalfont}{\thesection}{1em}{\MakeUppercase{#1}}
  \titleformat{\subsection}
  {\it}{\thesection}{1em}{{#1}}
  
\begin{document}
\small
\begin{strip}

\title{Theoretical limits of Virtual Reality}
\date{}

%% Authors and addresses/affiliations

\author{Francesco Sisini, Valentina Sisini, Laura Sisini\\
\begin{small}
  %Dipartimento di Fisica e Scienze della Terra, Universit\'a di Ferrara Ferrara, Via Saragat 1, 44122 Ferrara, Italy
  Tekamed srl
\end{small}\\\\
}

%\address[Affil2]{Vascular Diseases Center, University of Ferrara, Via Aldo Moro 8, 44124 Cona (FE), Italy}
% Replace capitalized text with the appropriate information (use standard capitalization rules for your text, not all capitals.

%\begin{CJK}{UTF8}{min}\Large 拳必殺  notes series\end{CJK}

\begin{center}
\line(1,0){450}
\end{center}
\maketitle

%% Do not remove the page break here.
\pagebreak

%\linenumbers

%Ikken hissatsu (\begin{CJK}{UTF8}{min}拳必殺 )\end{CJK} means something like \textit{to annihilate at one blow}.
%This document is part of a series of   notes each one targeting a single goal. Each note has  to \textit{annihilate  at one blow}!

\section*{Abstract}
In recent years there has been a strong development of the concept of virtual reality (VR) which from the first video games developed in the 60s has reached the current immersive systems.
VR and the consequent deception of perception pose an interesting question: is it possible to deceive the mind to the point of not being able to recognize whether the perceived reality is real or simulated? In addition to this question, another question arises spontaneously: is it possible to simulate a \textit{non-reality} in which the physical laws do not apply?\\
The answer to the first question is that it would theoretically be possible to deceive the mind to the point of not being able to recognize whether the perceived reality is real or simulated, furthermore it is also possible to simulate a non-real, i.e. magical, world. However, this possibility is based on hiding degrees of freedom from the observer, therefore it requires the complicity of the observer to be realised.\\
It can be said that a VR system can simulate both a real and \textit{non-real} experience.\\
For an observer, a virtual reality experience is still an experience of reality. This experience can be exchanged for a different experience, that is, for a different reality that can be real or not real.
\end{strip}

\section{Introduction}
In recent years there has been a strong development of the concept of virtual reality (VR) which from the first video games developed in the 60s has reached the current immersive systems.
The human being perceives the surrounding reality through the senses. To date, the known senses are sight, hearing, touch, taste and smell. The perception of reality can be deceived in particular situations in which one is induced to misclassify what is perceived with the senses, for example two parallel segments can be perceived as curved lines if they are superimposed on a bundle of straight lines with an origin close to them .
The view is deceived by what we call optical illusions, but in other cases it is not the view that is deceived, but our judgment as in the well-known "Fata Morgana" effect. Similarly to what happens for the "Fata Morgana", VR technologies are focused not on the deception of sight but rather on the deception of judgment.
The strategy of deception of the VRs is articulated in two different actions: the first consists in the obscuring of the external reality and the second in the administration of an alternative reality. Similarly to sight, the other senses can also be deceived by VR. Current technologies are mainly focused on sight, hearing and touch, while those for smell and taste, however important, have so far been only marginally developed.\\
VR and the consequent deception of perception pose an interesting question: is it possible to deceive the mind to the point of not being able to recognize whether the perceived reality is real or simulated? In addition to this question, another question arises spontaneously: is it possible to simulate a reality in which the physical laws do not apply? In this research we have given a definition of reality simulation and we have established the theoretical limits that limit the possibilities of VR.

\subsection{Relationship between physical systems and computation}
Consider an ordinary computer during a computation, i.e. running an application. From a physical point of view what happens in the computer is the dynamic evolution of an electronic circuit, i.e. a flow of electric charges which occurs in accordance with the laws of physics which describe the forces acting on the system\footnote {The initial conditions of the system can be seen as the application program}.

Any computation is therefore a physical process which can be described in formal terms by means of a set of variables representing the degrees of freedom of the system in question.

This vision of computation understood as dynamic evolution has been well described by David Deutsch in a seminal paper on quantum computing\cite{Deutsch}. Several authors have accepted this point of view (computation is a physical process) and have further developed it by posing the inverse question: for which physical systems and under which conditions can it be considered that the dynamic evolution (such as the oscillation of a pendulum\cite{Sisini}) represents a computation?\cite{Horsman}.
We will make use of this point of view, more physical than mathematical, to analyze the problem posed in this paper and establish what are the physical limits for VR. In particular, we will exploit the description of the physical computing system through its degrees of freedom.
In general, a computing system is a physical system with $N$ degrees of freedom, where $N$ is the cardinality of the set obtained by merging the set of degrees of freedom used for the input with those used for output and those used for computation only. Computation is therefore the sequence of states through which the system passes during its dynamic evolution.

\subsection{Degrees of freedom of a physical system}
\label{trajectories}
Physical systems, both material (particles, strings, planets, etc.) and immaterial (light, etc.), are described mathematically through the concept of degree of freedom (dof). The number of degrees of freedom of a system is the number of variables required to fully describe its physical state.
If the system has no constraints, the degrees of freedom of the system coincide with the coordinates of all the components of the system\footnote{Even quantum concepts such as spin are described by a degree of freedom. Similarly fields, both classical and quantum, are described by degrees of freedom.}
Otherwise, if there are relations between the components of the system which constrain the values that the degrees of freedom can assume (constraints) then the degrees of freedom are less than the coordinates of the system and all these can be obtained as functions of the degrees of freedom themselves. \\
If we consider a system composed of a certain number $n$ of particles free from constraints (considered dimensionless) and integrating with each other, then the number of degrees of freedom will be $3\times n$. Once the equations of dynamics have been solved, it is noted that the motion of each particle occurs along a curve and is parameterized by time. Thus the motion of each particle can be described on this curve using only one parameter (time). Consider, for example, a particle in a centrally symmetrical force field. For suitable initial conditions, said particle will move on a circular trajectory. Therefore, although the degrees of freedom of the system are 3, its motion can be described using a single parameter which describes the progress along the curve corresponding to the trajectory.
In practice, during the evolution of a physical system, the number of free parameters necessary to describe the motion is reduced to one, ie the passage of time, while the various components of the system move along trajectories determined by the equations of dynamics.
This aspect is crucial in the practical implementation of simulations for VR, so it is good to keep it in the right consideration from the beginning.
(image of the parabolic trajectory)

\subsection{Simulation process}
In common parlance, two possible meanings of the term simulation can be distinguished. The first refers to the use of numerical methods to evaluate the equations of a physical system at a given instant of time. An example is the simulation of atmospheric events that allows weather forecasts.

The second instead refers to a computer system that has the purpose of creating a deception of human perception in order to simulate a different reality from the one that would be perceived in the absence of the simulation. The result of the simulation is the instant by instant production of a virtual reality, which is transmitted to a person through specific tools that act on his senses. An example is video games.

These two types of simulation can also be present in a single process, as is the case with flight simulators used for pilot training.

\section{Conceptual non-interactive simulation systems}
\subsection{Dynamic evolution of a physical system}
Let us consider a builder who designs and assembles simple physical systems that we will call computers (progressively computer $\mathcal{A}$, $\mathcal{B}$, ...) and an observer who observes these computers. The builder shows his calculators to the observer.
The builder agrees with the observer which degrees of freedom of the system constitute the input, which the output and which are to be considered computational. The output degrees of freedom are those that the observer can observe.\footnote{An example could be the agreement between the oculus manufacturer and the wearer, to wear them in the correct way to view the headset output.}

The observer is equipped with a Cartesian reference system and a clock, so he can measure and keep track of what he observes.
\begin{itemize}
\item The measuring instruments of the observer have a finite and non-zero sensitivity.
\ item The builder declares to the observer that the computer he built simulates a specific physical system.
\end{itemize}
The observer observes the calculator and measures for a certain period of time the values assumed by the variables that represent its output.
Using the measurements obtained, the observer draws a curve parameterized by time which represents the trajectory of the representative point of the observed system. It is possible that there exists a set of differential equations which admits such a curve as a solution. If this system of equations is defined in the same set of variables (or in a set obtained through an invertible transformation) that is used to describe the output of the system, then the observer will declare that the observed system is a real physical system\cite {epr}. Furthermore, if the system of equations found coincides with the equations describing the physical system that the builder claims to simulate, then the observer will agree with the builder.

If the observer does not find this system of equations, he could resort to one of the following alternatives to bring his own curve back to a system of differential equations:
\begin{itemize}
\item Add dynamic variables to equations that are not present in the calculator output
\item Add rheonomic constraints whose shape could also change between two different observations
\end{itemize}
In both these cases the observer will declare that the system simulated by the computer is a non-real physical system and will not agree with the manufacturer.
Below we consider a series of calculators made by the manufacturer and see which of them the observer agrees with the manufacturer.

\subsection{Computer $\mathcal{A}$}
Let us consider a calculator $\mathcal{A}$ composed of a box with very massive walls and a bottom in which a disk can slide without friction.

The physical system of the $\mathcal{A}$ calculator can be described using only two degrees of freedom and limiting the dynamics to the space enclosed by the edges of the box.
The $\mathcal{A}$ calculator can be thought of as a computational system whose degrees of freedom are the $x$ and $y$ coordinates of the disk. An observer can observe the disk moving and bouncing against the edges of the box and can measure the coordinates of the disk at any instant of time.
The manufacturer declares that the system simulates the free motion of a disk in a square space with side $l$. For the $\mathcal{A}$ calculator, the constructor establishes that there are no input degrees of freedom and that the diskette coordinates $x$ and $y$ constitute the system output. Once this agreement is accepted by the observer he agrees that from his observation, the $\mathcal{A}$ calculator is effectively simulating a physical system in which a disk moves with inertial motion in a space bounded by four constraints of the type $x<l/2$, $x>-l/2$, $y<l/2$ and $y>-l/2$ as there is a set of differential equations which, together with four scleronomic constraints, describes what has been observed without the need to introduce hidden variables or further constraints in addition to those declared by the constructor.

\begin{figure}
  \includegraphics[width=0.5\textwidth]{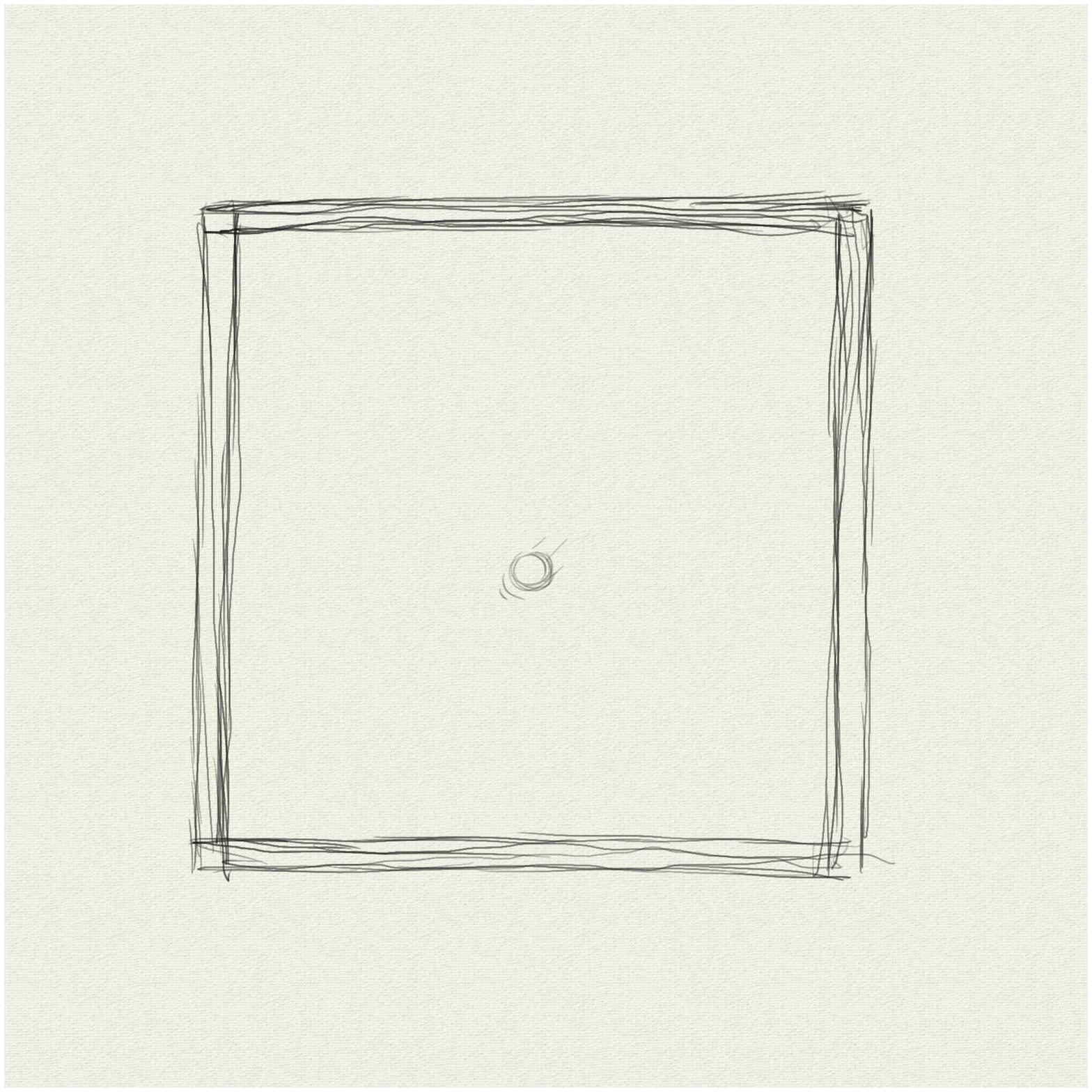}
  \caption{{\bf Computer $\mathcal{A}$}
    A disk of mass $m$ is free to move inertially with little friction in a container with very massive walls.
  }
  \label{fig:c_A}
 \end{figure}

 \subsection {Computer $\mathcal{B}$}
Let's consider a second calculator that we will call $\mathcal{B}$, similar to the $\mathcal{A}$ calculator but which contains two diskettes instead of just one.

If the set $\mathbf{O}$ of the output degrees of freedom declared by the constructor corresponds to the coordinates $x_1,\ x_2,\ y_1$ and $y_2$, then the same considerations made for the $\mathcal{ calculator are valid A}$, ie the observer agrees that the $\mathcal{B}$ calculator is simulating a physical system consisting of two free-moving disks. However, the constructor declares that for the $\mathcal{B}$ calculator the degrees of freedom dedicated to the output are defined by the set $\mathbf{O}=\{x_1,y_1\}$ therefore the output is limited to degrees of freedom $x_1$ and $y_1$, i.e. to the first disk only, while the second disk has a computational role only, and is part of the set $\mathbf{C}=\{x_2,y_2\}$ of the degrees system computations.
To get a simple mental picture of such a calculator, imagine that you are looking at the $\mathcal{B}$ calculator in the dark and that you can only see one disk because a light source has been placed on it.

In the dark, the observer will only be able to see the luminous disk and will see it proceed in inertial motion until it meets the second disk, then will see it change trajectory abruptly. Not being able to see the second disk, he will not be able to identify a system of differential equations whose solution is the trajectory of the motion of the luminous disk without introducing a set of hidden variables or rheonomic constraints. It should also be noted that for each possible computation of the computer, the manufacturer could modify the initial conditions which fix the position and the initial speed of the disks. Under these conditions, the additional constraints that the observer must impose are not only rheonomic but also arbitrary, since their functional form can vary from time to time.

\begin{figure}
  \includegraphics[width=0.5\textwidth]{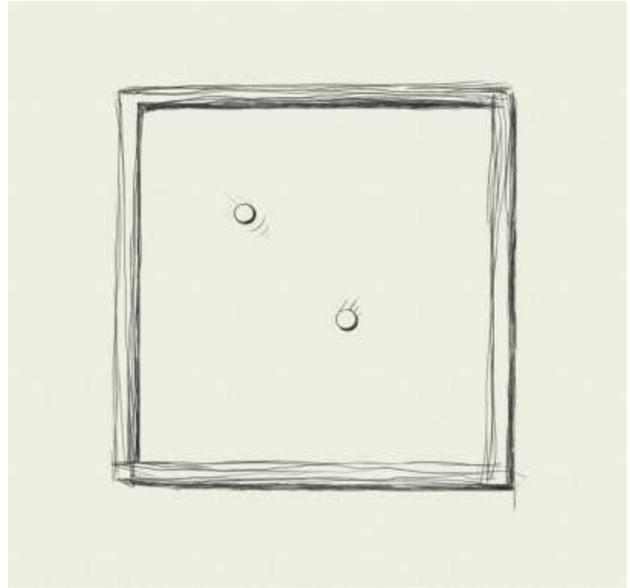}
  \caption{{\bf Computer $\mathcal{B}$}
Two diskettes, both of mass $m$, are free to move by inertial motion with little friction in a container provided with very massive walls.
  }
  \label{fig:c_B}
 \end{figure}

\subsection{Computer $\mathcal{C}$}
Let us consider a third calculator $\mathcal{C}$ in which there is only one disk but this is connected to two opposite walls with two springs invisible to the observer. The manufacturer declares that the system output consists of the disk which is the only visible detail of the system. The manufacturer declares that the system is simulating a harmonic motion. The observer agrees with what the manufacturer affirmed, in fact he is able to identify a system of differential equations whose solution represents the trajectory of the motion of the observed disk.

\begin{figure}
  \includegraphics[width=0.5\textwidth]{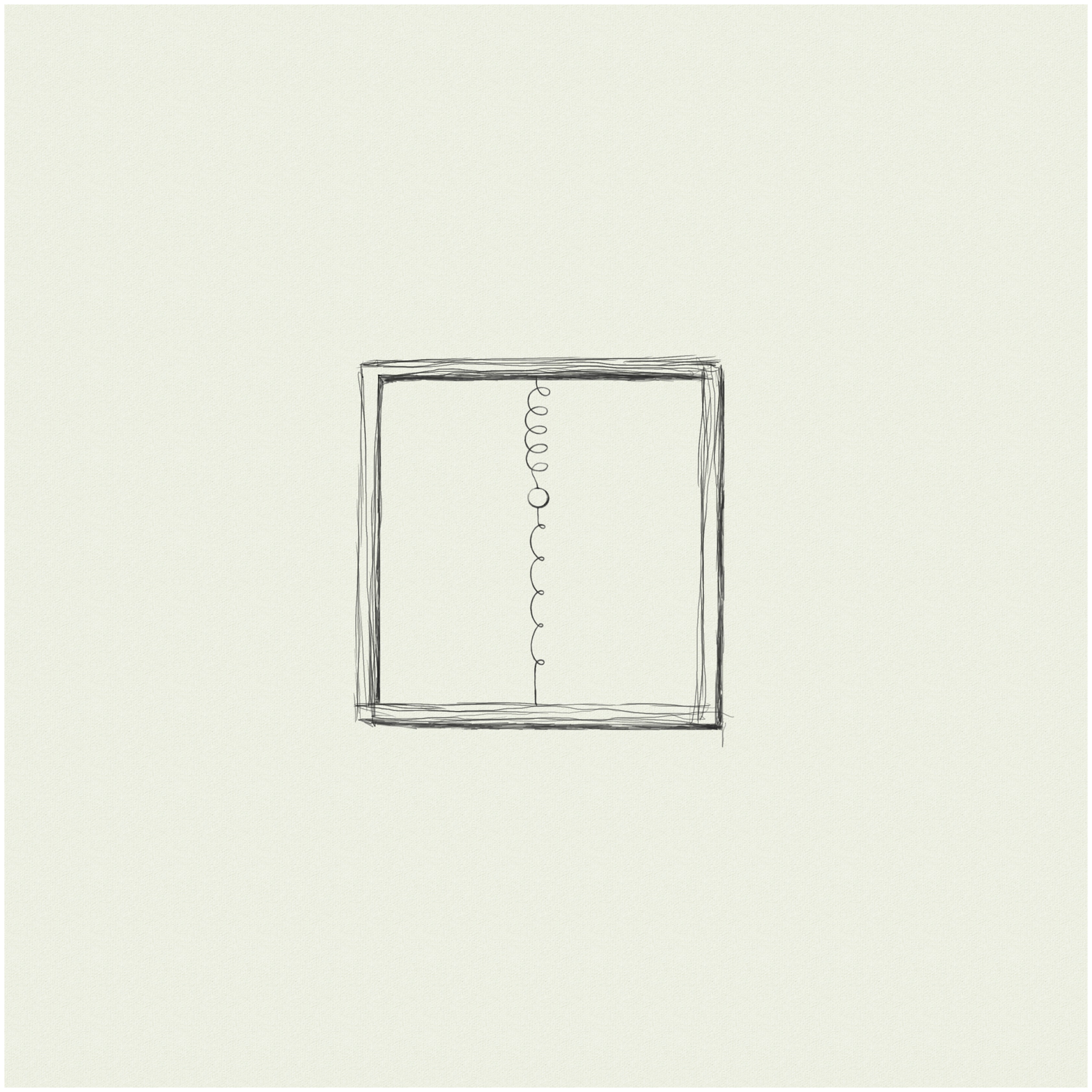}
  \caption{{\bf Computer $\mathcal{C}$}
A disk of mass $m$ is connected to two springs which are connected to the two opposite faces of a container. The diskette moves by harmonic motion with little friction in a container provided with very massive walls.
  }
  \label{fig:c_C}
 \end{figure}

\subsection{Computer $\mathcal{D}$}
The $\mathcal{D}$ calculator is obtained from the $\mathcal{C}$ by inserting a dividing wall in the box and a second disk free to move by inertial motion in the half of the box in which the motion of the first disk does not occur. So adding two computational degrees of freedom not visible to the observer.
The manufacturer declares that the system is simulating a harmonic motion. The observer again agrees with what the manufacturer said, in fact he is able to identify a system of differential equations whose solution represents the trajectory of the motion of the observed disk without being disturbed by the second disk which does not interact with what he observed.

\begin{figure}
  \includegraphics[width=0.5\textwidth]{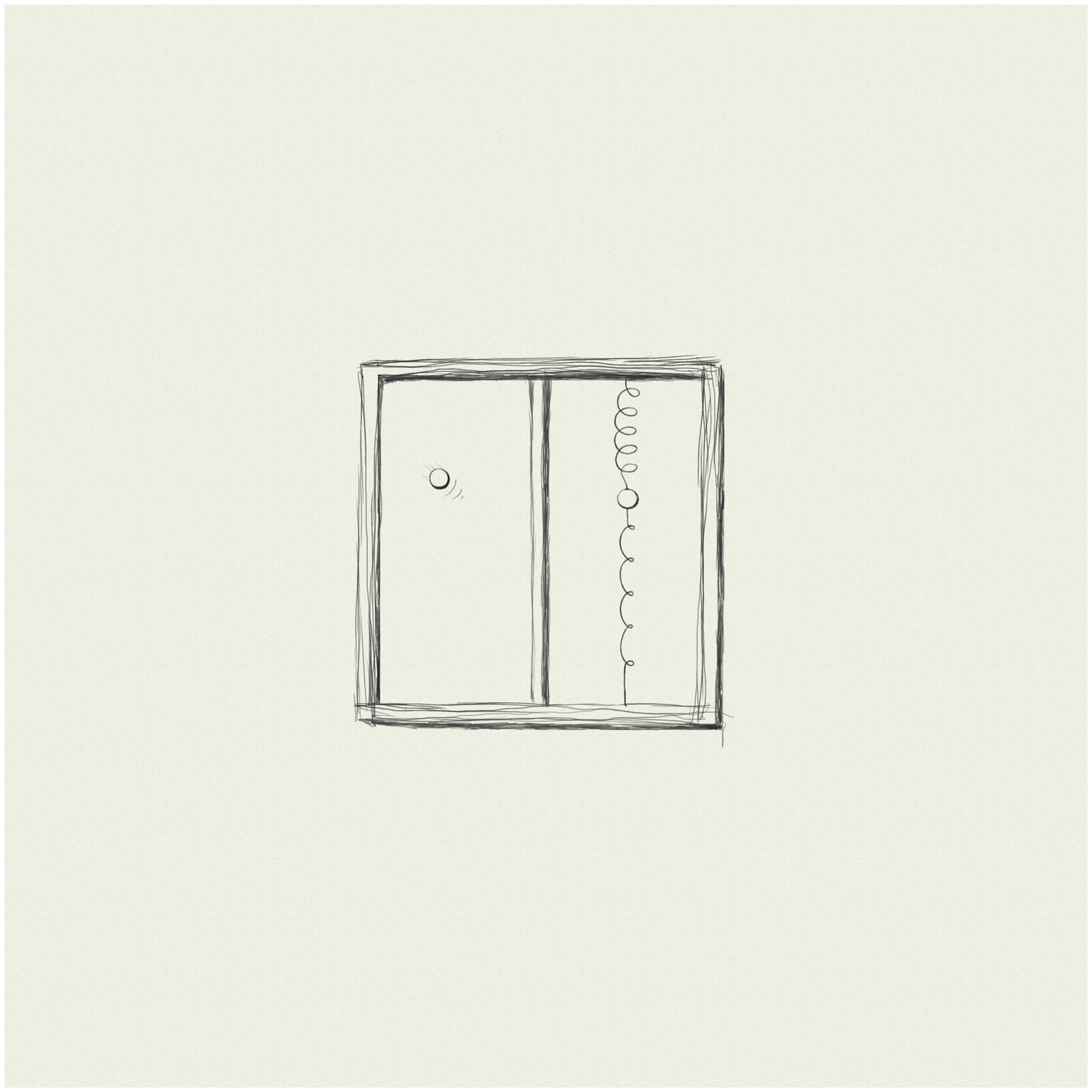}
  \caption{{\bf Calcolatore $\mathcal{D}$}
A diskette is free to move by inertial motion inside a container while another diskette moves by harmonic motion alongside it. The two disks are kept separate by a wall between them.
  }
  \label{fig:c_D}
 \end{figure}

\subsection{Computer $\mathcal{E}$}
This calculator consists of a disc hinged at its center and free to rotate. In one point of the disk a light source is applied and it is identified with the output of the system, while the rest of the disk is invisible to the observer.
The manufacturer declares that the system is simulating a motion due to a centrally symmetrical force field with module $F$ described by a certain constant $k$ and by the charge $a$ placed at a distance a $r$ from the center of symmetry:
\begin{equation}
  F=k\frac{a^2}{r^{2}_{12}}
\end{equation}

Using the measurements made on the output of the calculator, the observer traces the curve which represents the trajectory of the representative point of the system and observes that it is a circumference parametrized by time by means of a relation which proportionally links time to the subtended angle from puto with a fixed axis. This curve represents the solution of the differential equations of the system described by the manufacturer, therefore the observer agrees with the manufacturer. It is noted that with this calculator for the first time the manufacturer seeks agreement with the observer by declaring that he simulates a different reality from the one he is actually calculating.

\begin{figure}
  \includegraphics[width=0.5\textwidth]{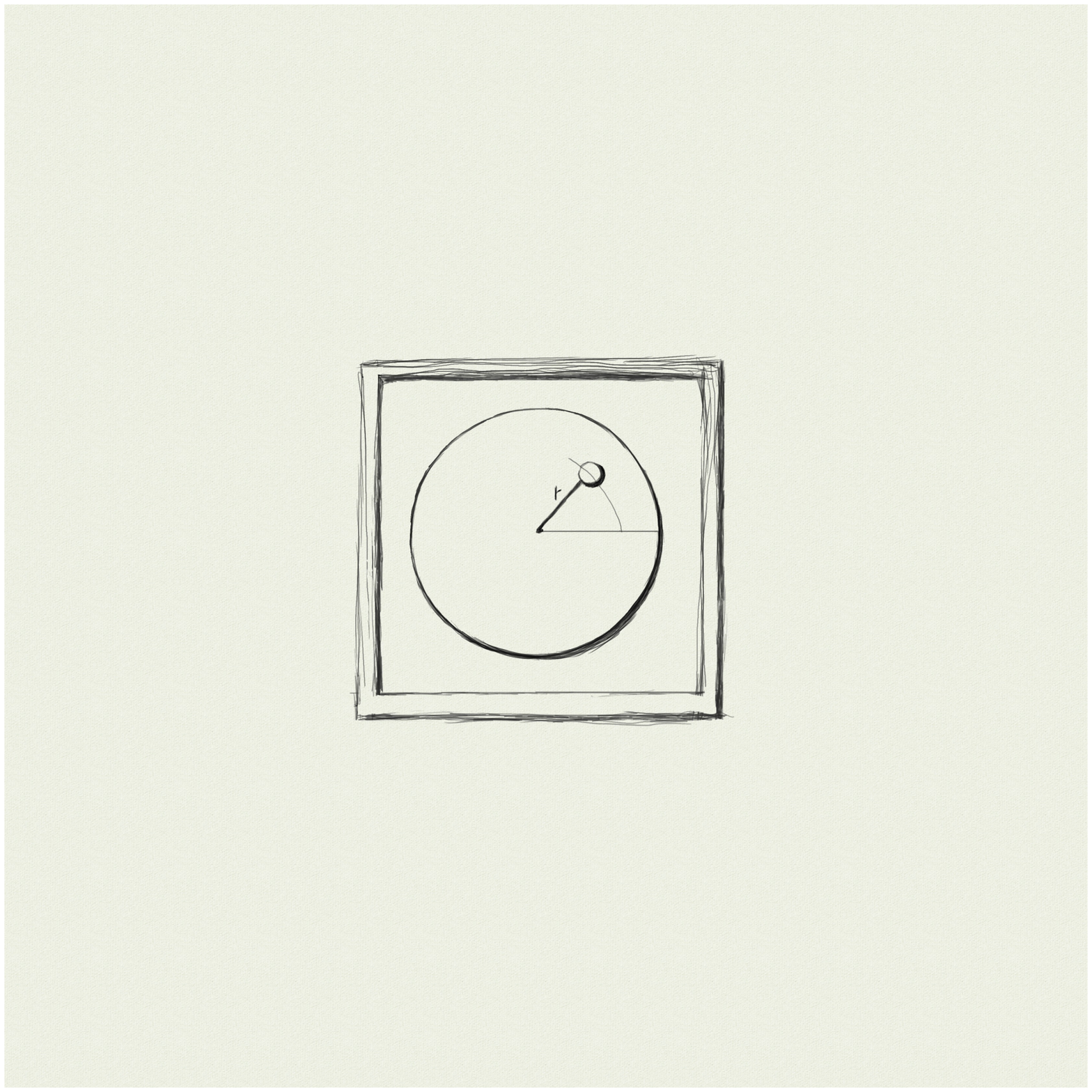}
  \caption{{\bf Computer $\mathcal{E}$}
A disk of mass $m$ is hinged at its center to a pivot so that it can rotate about its center of mass. A light source at a distance $r$ from the center is applied to it. The mass of the light source is considered to be negligible compared to the mass of the disk.
  }
  \label{fig:c_E}
 \end{figure}

\section*{Interactive conceptual systems, i.e. Virtual Reality (VR)}
The calculators seen in the previous section expect that the observer is limited to observing the output of the simulation. In practice, the sensory activity of the observer is reduced to passive perception only: he cannot touch.\\
Now we introduce a new type of calculator that we will call VR which provides for a greater participation of the observer. This type of calculator also provides input degrees of freedom that we will denote with the set $\mathbf{I}$. With this calculator the observer is not limited to observing the output but can act with forces on the input degrees of freedom and therefore he becomes active and becomes part of the computation system, i.e. of the calculator itself.
\subsection{Computer $\mathcal{F}$}
This calculator is built starting from the $\mathcal{A}$ model, but a rod with mass $m_c<<m_d$ of the disk is applied to the disk. This rod constitutes the controller with which the observer can apply an arbitrary force to the puck. Acting on the disk with the controller, he verifies that the disk accelerates according to Newton's second law and therefore again agrees with the manufacturer's statement that the $\mathcal{F}$ calculator represents a VR of a free disk.
\begin{figure}
  \includegraphics[width=0.5\textwidth]{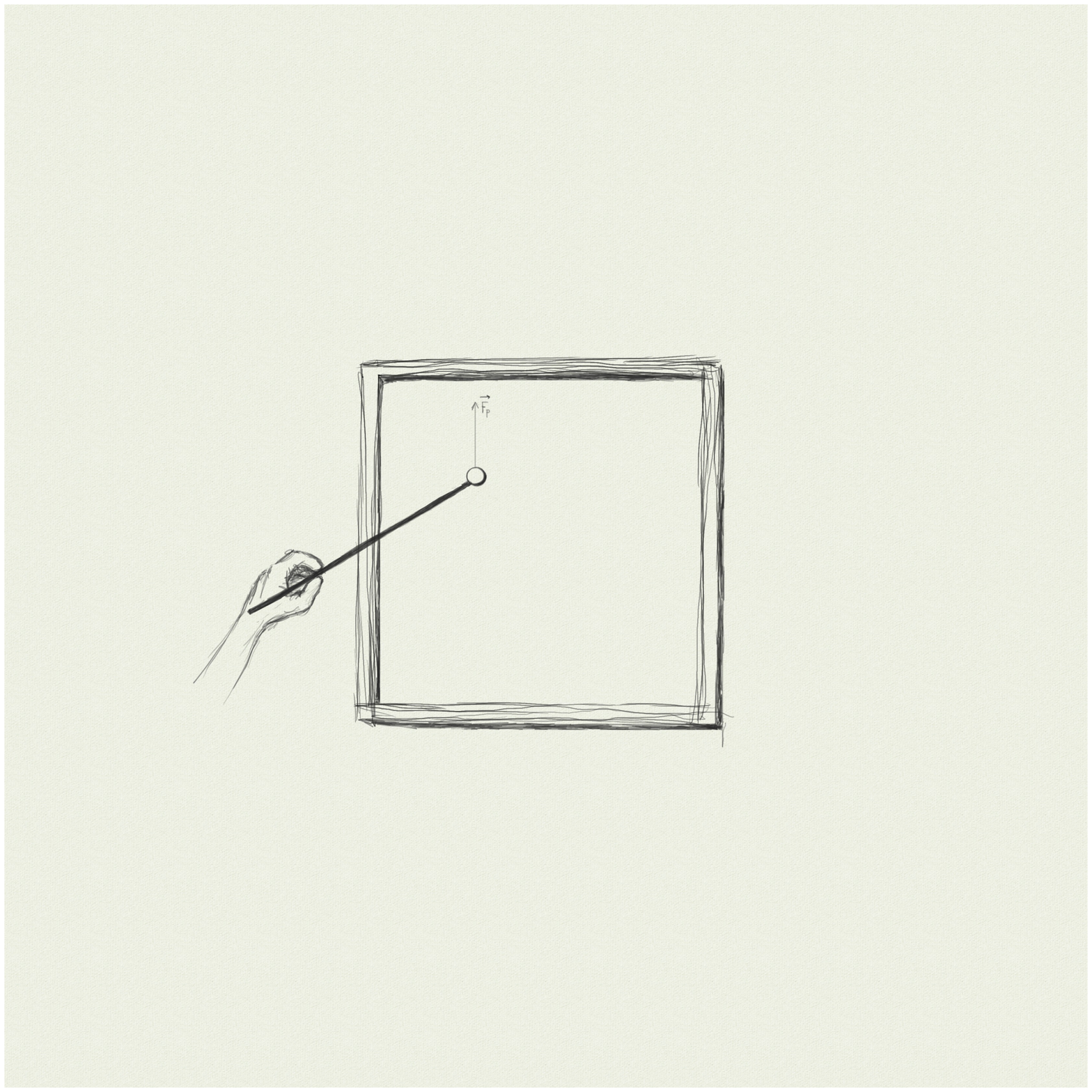}
  \caption{
      A disk of mass $m$ is free to move inertially with little friction in a container with very massive walls. A small rod is hinged to the disk which allows a force to be applied to it.
  }
  \label{fig:c_F}
 \end{figure}

\subsection{Computer $\mathcal{G}$}
This calculator is built starting from the $\mathcal{B}$ model, where now only one of the two disks present has a control rod applied.
As happens for the $\mathcal{B}$ calculator, also with the $\mathcal{G}$ calculator the observer does not agree with the VR proclaimed by the manufacturer.
\begin{figure}
  \includegraphics[width=0.5\textwidth]{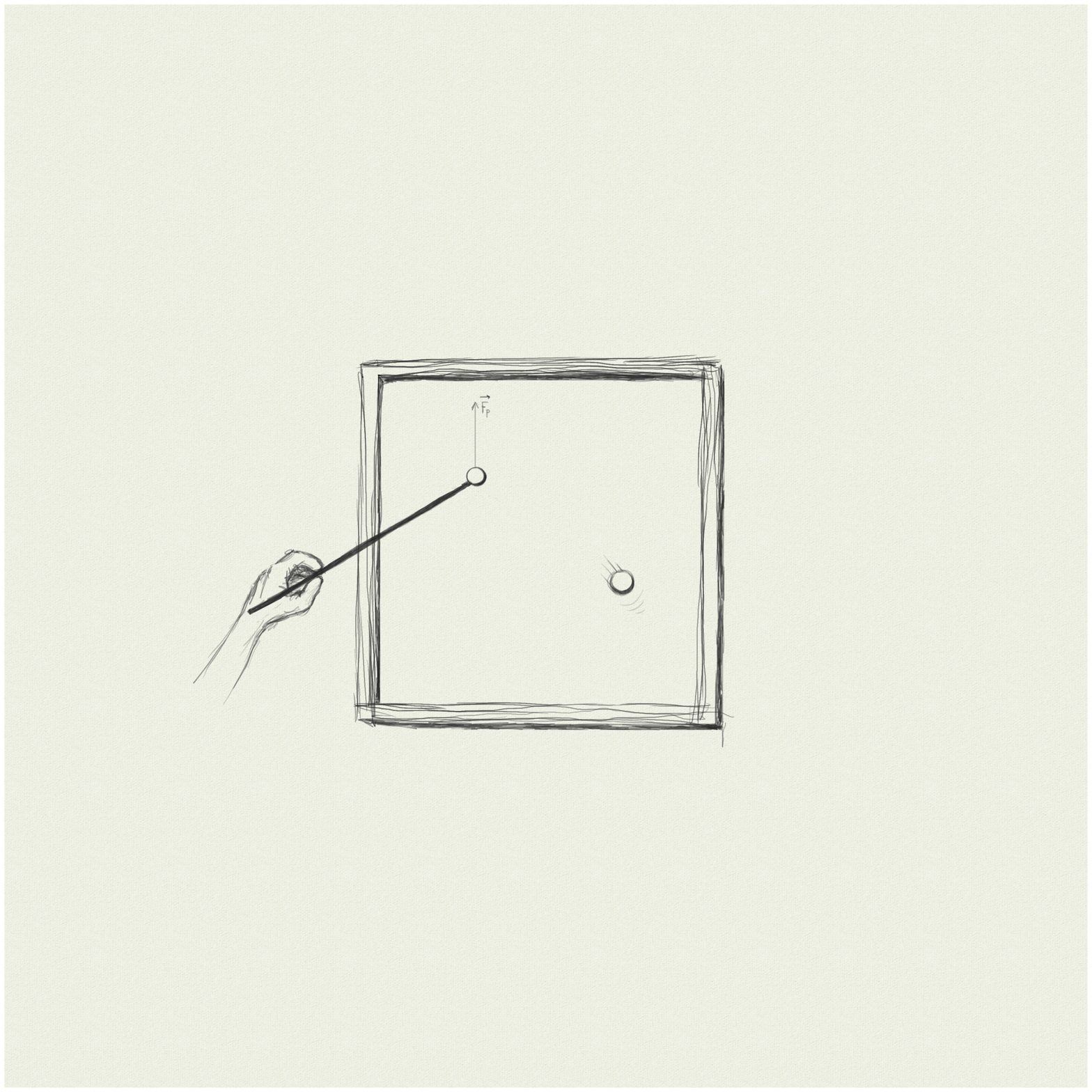}
  \caption{{\bf Computer $\mathcal{G}$}
      A disk of mass $m$ is free to move inertially with little friction in a container with very massive walls. A small rod is hinged to the disk which allows a force to be applied to it. A second puck is free to move without the action of external forces.
  }
  \label{fig:c_G}
 \end{figure}

\subsection{Computer $\mathcal{H}$}
The $\mathcal{H}$ calculator is obtained by applying a controller to the luminous point $p$ of the rotating disk presented in the calculator E. By applying a force to the point $p$ the observer verifies that the observed system is not described by a potential with central symmetry as declared by the manufacturer, in fact he verifies that he can stop the motion of the point $p$ without its trajectory collapsing towards the center as it would happen if it were the declared physical system.
\begin{figure}
  \includegraphics[width=0.5\textwidth]{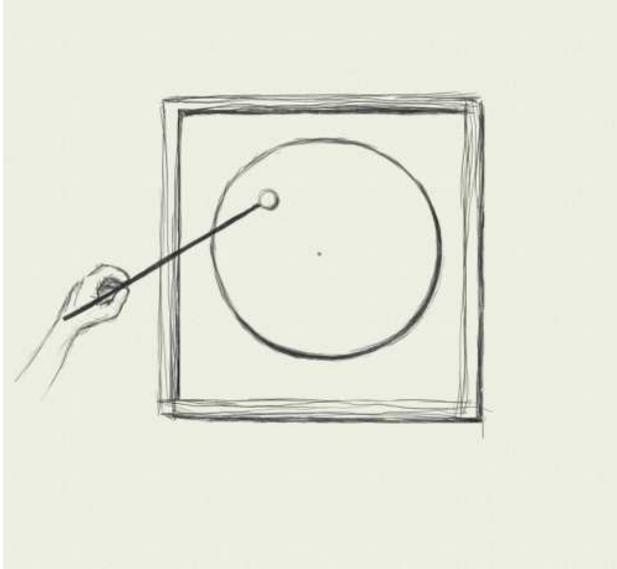}
  \caption{{\bf Computer $\mathcal{H}$}
   A disk of mass $m$ is hinged at its center to a pivot so that it can rotate about its center of mass. A light source at a distance $r$ from the center is applied to it. The mass of the light source is considered to be negligible compared to the mass of the disk. A small rod is hinged in correspondence with the light source which allows you to apply a force at will.
  }
  \label{fig:c_H}
 \end{figure}

\subsection{Computer    $\mathcal{X}$}
This calculator consists of a toothed linear element which alternately comes into contact with two toothed wheels which rotate by inertia, one clockwise and the other anti-clockwise. The moment of inertia of the two wheels is assumed to be much greater than that of the linear element which, as a first approximation, is considered as having mass equal to zero. Each wheel is hung on a spring. Between the two motions there is a phase shift such that they are not found to act simultaneously on the linear element. The result is that the linear element moves along the axis $x$ with uniform rectilinear motion in an alternating direction. On this element there is a light source which is considered the output of the system. The manufacturer declares that the system has two degrees of freedom, i.e. the coordinates $x$ and $y$ of the light source and also declares that it moves by inertial motion.

\begin{figure}
  \includegraphics[width=0.5\textwidth]{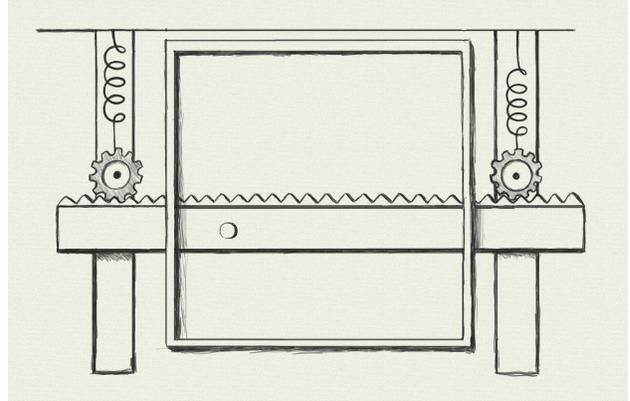}
  \caption{{\bf Computer    $\mathcal{X}$}
     A linear toothed element is scrolled horizontally through a gear system that controls its movement. A light source is placed on this element.
  }
  \label{fig:c_G}
 \end{figure}

\section{Theoretical limits of conceptual interactive systems (VR)}
In the previous sections we have seen that a computing system (computer) is a physical system described by a certain set $\mathbf{C}$ of degrees of freedom. Similarly to the computation system, the output subsystem is also described by a certain set $\mathbf{O}\subseteq\mathbf{C}$ of degrees of freedom and the same is true for the input subsystem $\mathbf{I}\ subseteq\mathbf{C}$. Having made these considerations, it is possible to draw considerations of a general nature which we see below.

\subsection{Simulation systems without input devices}
These systems can simulate both a physical system (calculator $\mathcal{A}$) and a non-physical system (calculator $\mathcal{B}$), i.e. a system that is not describable through observable degrees of freedom but is possibly traceable to a physical description if we consider the presence of hidden variables. \\

It should be noted that if the observer can observe all the degrees of freedom of the system, i.e. $\mathbf{O}\equiv \mathbf{C}$ then he always identifies the computer with a physical system and agrees with what is declared by the manufacturer only if the constructor declares a physical system compatible with the equations obtained by the observer.\\

Instead, if the system has computational degrees of freedom that are not observable $\mathbf{O}\subset \mathbf{C}$ the observer's judgment depends on the relationship existing between the computational degrees of freedom. If the unobservable degrees of freedom $\mathbf{U}$ $(\mathbf{U}=\mathbf{C}-\mathbf{O})$ do not interact with the observable ones $(\mathbf{O})$ the system will always be classified as physical by the observer (calculator $\mathcal{D}$). Otherwise, if the degrees of freedom $\mathbf{U}$ interact with the degrees of freedom $\mathbf{O}$ then the observer will classify the non-physical system because, as already observed, he will not be able to deduce a system of equations based only on the variables $x_i\in \mathbf{O}$ (calculator $\mathcal{B}$).

It should be noted that even in the case where $\mathbf{O}\equiv \mathbf{C}$ it can happen that the manufacturer claims to simulate a physical system different from the one actually implemented, so in these cases the observer agrees with a reality wrong physics, but still plausible (calculator $\mathcal{E}$).
With reference to what has been said in the \ref{trajectories} section, this likelihood is generated because the computer is not analogous to the system it is simulating but, in a condition of inertial motion, each point on it follows the same trajectory followed by the representative point of the system declared by the manufacturer (particle in a field with central symmetry). Therefore: the system declared by the manufacturer foresees two degrees of freedom but admits uniform circular motion as a solution. The system actually implemented by the manufacturer is a system with a single degree of freedom which in inertial motion moves with uniform circular motion. 

\subsection{Simulation systems with input devices}
As we have seen, the presence of controllers in computers gives the observer the ability to act on the computer input degrees of freedom $\mathbf{I}$ by applying a force to it. This ability gives him yet another tool for classifying the underlying physical system.
In case the controller can act on all $\mathbf{C}$ degrees of freedom of the system, then the observer will always be able to classify the system correctly, as happens with the $\mathcal{H}$ calculator when the observer realizes that the system is different from what was declared by the manufacturer. While otherwise, the observer will keep the same dilemmas.

\section{Concrete systems of Virtual Reality}
The calculators that have been described in the previous paragraphs are based on the mechanics of the material point (or of the rigid body). In the real world, VR is produced by computer systems that are not based on the mechanics of the material point that we have used in this dissertation, but on digital electronics, i.e. on the physics of classical electromagnetism\footnote{In reality, the LED devices used to make video screens they require both mechanical explanations, but the same result could be obtained using a cathode ray tube instead of LEDs, so that we can say that current VR systems can already be easily explained with classical electromagnetism.}.
However, they are physical systems whose dynamics can be described by a certain number of degrees of freedom which, as for mechanical computers, can be divided into degrees of input, output and computation.
So even for modern VR systems we can say that the same considerations elaborated for mechanical computers that we have dealt with up to now are valid.

In real examples, the output degrees of freedom are in the pixels projected by the oculus and the sound frequencies produced by the earphones.

As an example, in the first proposed example of the floppy disk bouncing in a box (computer $\mathcal{A}$) it can be described as an ordinary computer connected to a screen. In a nutshell, the screen is an electronic device to which a matrix of LEDs is connected. A few-instruction software is running on the computer. Each instruction is encoded as an electrical charge stored in a capacitor array. The disk is displayed by executing the code which calculates its coordinates and transforms them into physical addresses of the LEDs connected to the screen. Code execution is governed by computer CPU cycles: electrical oscillations. The observer visualizes the disk, that is a subset of the LEDs activated on the basis of what is computed by the CPU. As you can easily understand, even what is perceived as an image on a screen is still the set of a certain number of degrees of freedom of a physical system, those defined for the output. So even in the electronic version of the bouncing disk, the observer is observing part of the degrees of freedom of an evolving dynamic system.

Having made this necessary premise to link the conceptual reasoning set out above to practical considerations involving current systems, let us come to the discussion of the results.

\section{Discussione}
The results presented in the previous section allow us to outline a picture of the theoretical limits of the possibilities offered by VR.

Starting from the various cases exposed, one can reason inductively and affirm that a physical system of $N$ degrees of freedom can be simulated using a physical system with a number $M$ degrees of freedom which can be greater, less than or equal to $N$.
In more direct terms, wearing oculus and earphones, one can witness a simulated reality, i.e. one can observe a reality that is not happening where we believe but is taking place in the computer in the form of a physical process that passes through a set of states similar to that of perceived reality.

In practice, therefore, a computerized system can actually simulate a physical system, that is, it can mislead an observer into agreeing on the likelihood of what he is observing with respect to what is declared by the producer of the simulation.\\
However, the simulation can take place in different ways:
\begin{enumerate}
\item the system to be simulated consists of $N$ degrees of freedom and is simulated using a calculator with $P$ degrees of freedom of which $M$ degrees of freedom output with $M<N$. In this case the simulating calculator is not physically analogous to the system to be simulated but its output can be (calculator $\mathcal{E}$).
\item The system to be simulated consists of $N$ degrees of freedom and is simulated using a calculator with $N$ degrees of freedom all labeled as output degrees. In this case the simulating calculator coincides with the system to be simulated (calculators $\mathcal{A}$ and $\mathcal{C}$).

\item the system to be simulated consists of $N$ degrees of freedom and is simulated by a computer with $M>N$ degrees of freedom of which $N$ are used as output. If the $N$ of $M$ output degrees of freedom are physically independent of the remaining $M-N$ degrees of freedom (calculator $\mathcal{D}$), then we return to the previous case, otherwise this means that the degrees observable degrees of freedom (ouput) are dependent on hidden degrees of freedom ($\mathcal{B}$ and $\mathcal{X}$ calculators).
\end{enumerate}
In the first simulation mode we see that a computer with limited resources can simulate a more complex reality as it does not simulate the interaction between the system components but only the trajectories of the system components which generally take place on parameterized curves. This allows for example to transform a system with two degrees of freedom (point subject to a central potential) into a system with only one degree of freedom (point on a rotating circle). This type of simulation remains plausible only if the observer simply observes.
If instead the observer has an input system towards the computer (ie a controller) then he realizes that the reality he has joined does not correspond to the physics of the system, therefore the simulation does not hold up to experimental verification. \\  
In the second way the computer and the system to be simulated coincide. In this case the simulation is a ``staging'' and is absolutely resistant to the action of the observer who can act with the controller on all the degrees of freedom of the system without running into any inconsistencies. The limit of this way of simulating is evident in that to simulate a given reality it is necessary to create it in all its details by transforming the simulation into a copy of reality

The third way is the most used and consists in using more degrees of freedom than the observable ones in the simulation. This mode allows to simulate reality as long as the observer limits himself to observing, but if the observer has a controller then he, acting on a given degree of freedom, will experience that it is linked to other hidden degrees of freedom, therefore he will the consistency of the simulation.

At this point we are able to formulate the answers to the initial questions.
The answer to the first question is affirmative: it is possible to deceive the mind to the point of not being able to recognize whether the perceived reality is real or simulated. However, this deception only holds up as long as the simulation is passively undergone but it does not hold up if it is possible for the observer to interact with the simulated reality.

We have also formulated an answer for the second question. In fact, just as it is possible to simulate reality, it is also possible to simulate a non-real, i.e. magical world, therefore: flying fairies, objects that remain suspended from the ground in defiance of the law of gravity, objects that appear out of nowhere or that vanish without conservation of mass etc. However, this possibility is based on hiding degrees of freedom from the observer, therefore it requires the complicity of the observer to be realised, exactly like the simulation of reality.\\
In conclusion it can be said that a VR system can simulate both reality and magic. To completely simulate reality, however, it is necessary to use more degrees of freedom than those describing the reality to be simulated. This is an important limitation because obviously there is a conceptual limit to this process.\\
In summary, for an observer, a virtual reality experience is still an experience of reality. This experience can be exchanged for a different experience, that is, for a different reality that can be real or not real.

\pagebreak

\end{document}